\documentclass[11pt,a4paper]{article}

\usepackage[T1]{fontenc}
\usepackage[utf8]{inputenc}
\usepackage{lmodern}
\usepackage{microtype}
\usepackage{amsmath,amssymb,amsthm,mathtools,bm,mathrsfs}
\usepackage{geometry}
\usepackage{booktabs}
\usepackage{longtable}
\usepackage{multirow}
\usepackage{array}
\usepackage{graphicx}
\usepackage{enumitem}
\usepackage[authoryear,round]{natbib}
\usepackage{hyperref}
\usepackage[nameinlink,noabbrev]{cleveref}
\usepackage{caption}
\usepackage{xcolor}
\graphicspath{{figures/}}
\hypersetup{
  hidelinks,
  pdftitle={Finite-Strength Sensitivity and Euler--UTSD Correspondence in Guderley--Mach Reflection},
  pdfauthor={Justin Kin Jun Hew},
  pdfkeywords={weak shock reflection, Guderley reflection, UTSD, discrete adjoint, full Euler, asymptotic-preserving limit}
}
\setlist{nosep}
\allowdisplaybreaks

\newtheorem{theorem}{Theorem}[section]

\theoremstyle{definition}
\newtheorem{assumption}[theorem]{Assumption}

\newcommand{\dd}{\mathrm{d}}
\newcommand{\OO}{\mathcal{O}}
\newcommand{\eps}{\varepsilon}
\newcommand{\al}{\alpha}
\newcommand{\del}{\delta}

\newcommand{\jump}[1]{\left[#1\right]}

\newcommand{\ip}[2]{\left\langle #1,#2\right\rangle}
\newcommand{\bU}{\bm U}
\newcommand{\bR}{\bm R}
\newcommand{\bPsi}{\bm\Psi}
\newcommand{\bx}{\bm X}
\newcommand{\PF}{\mathrm{PF}}
\newcommand{\Eul}{\mathrm{E}}

\title{Finite-Strength Sensitivity and Euler--UTSD Correspondence\\
for Guderley--Mach Reflection}
\author{Justin Kin Jun Hew\\
\small Australia's Climate Simulator (ACCESS-NRI)\\
\small Canberra, ACT 2601, Australia\\
\small \texttt{justinkinjun.hew@anu.edu.au}}
\date{}

\begin{document}
\maketitle

\begin{abstract}
Weak shock reflection at nearly glancing incidence is governed, after the
transonic weak-shock scaling, by a self-similar unsteady transonic
small-disturbance (UTSD) free-boundary problem.  We derive and differentiate
the first finite-strength perturbation of the corresponding isentropic
potential-flow problem along paths of fixed canonical incidence
$a=\alpha/\delta$, where $\mu=\delta^2=2(M^2-1)$.  A second-order refluxed
adaptive finite-volume method, exact discrete tangents and adjoints, and a
Rankine--Hugoniot-constrained fitted principal front give the fixed-$a$
canonical sensitivity
$H_{2,a}^{\PF}(0.5;1.4)=-0.217\pm0.012$; a fully differentiated physical
back-map gives the diagnostic fixed-$a$ coefficient
$K_{2,a}^{\PF}\simeq-0.266$.  We derive the exact chain rule that converts
these quantities to the distinguished fixed-$\lambda$ path
$\lambda=(M-1)/\alpha^2$, showing explicitly that the conversion requires the
independent incidence derivative of the leading UTSD branch and therefore
cannot be inferred from the fixed-$a$ calculation alone.  A matched-boundary
self-similar Euler study with strength-dependent refinement contains 21
qualified nonlinear states and 252 evaluations of a common front-functional
family.  Coupled extrapolation gives
$g_0^{\Eul}=0.510\pm0.006$ and the physical leading-angle coefficient
$G_0^{\Eul}=0.256\pm0.004$, consistent with the shock-fitted UTSD limits.
A separate same-strength phase-bracket audit using 18 Euler states shows that
the captured-shock subcell phase is comparable to the desired cubic signal.
The resulting finite-resolution Euler secants are compatible with the
potential-flow correction, but their $\rho=h_\eta/\sqrt\mu\to0$ extrapolation
is not model-stable.  Thus leading-order Euler--UTSD correspondence is
numerically verified, whereas cubic-order Euler correspondence remains
unresolved.
\end{abstract}

\noindent\textbf{Keywords:}
weak shock reflection; Guderley reflection; UTSD; potential flow; full Euler;
discrete adjoint; shock fitting; adaptive mesh refinement; asymptotic-preserving
limit.

\section{Introduction}
\label{sec:introduction}

The transition between regular and Mach reflection becomes singular when a
weak incident shock approaches a rigid wall at a nearly glancing angle.  In the
classical three-shock construction, the admissible deflection interval
collapses as the shock strength vanishes; experiments and high-resolution
computations instead reveal a weak or Guderley--von Neumann reflection with a
curved compression structure and a mixed subsonic--supersonic interaction
region \citep{ColellaHenderson1990,HunterBrio2000,SkewsAshworth2005}.  The
appropriate leading model is the two-dimensional UTSD system.  Its
self-similar Guderley--Mach branch contains shocks, expansion regions, sonic
boundaries, and successively smaller supersonic patches
\citep{HunterBrio2000,TesdallHunter2002,HunterTesdall2003}.

Define
\begin{equation}
 \del=\sqrt{2(M^2-1)},
 \qquad a=\frac{\al}{\del},
 \qquad \mu=\del^2,
 \label{eq:scales}
\end{equation}
where $M$ is the incident normal Mach number and $\al$ is the angle between the
incident shock front and the wall.  At fixed canonical incidence $a$, the
leading UTSD solution is independent of $\gamma$, but its first finite-strength
correction need not be.  For the fitted canonical attachment functional used
below, we write
\begin{equation}
 G_{\rm can}(a,\mu)
 =g(a)+\mu H_{2,a}(a;\gamma)+\OO(\mu^2),
 \label{eq:canonical-expansion}
\end{equation}
The quantity $g$ is not available in elementary closed form; it is a global
functional of the self-similar free-boundary solution.  The same is true of $H_{2,a}$.  The subscript records that the derivative is
taken at fixed canonical incidence $a$.

The present paper addresses two logically distinct questions.  First, can the
finite-strength derivative of the canonical potential-flow branch be defined,
computed, and numerically qualified?  Second, does the leading full-Euler weak
limit approach the same UTSD branch when grid and shock-strength limits are
taken together?  Treating these questions separately is essential.  Exact
differentiation of a discrete potential-flow calculation does not by itself
establish physical convergence, while an Euler calculation on a fixed grid
need not approximate the singular weak-oblique-shock limit.

The main contributions are:
\begin{enumerate}[label=(\roman*)]
 \item a conservative finite-strength potential-flow residual, including the
       interior forcing, incident-front motion, and corrected post-shock state;
 \item a continuous free-boundary linearisation and objective-adjoint identity,
       together with exact discrete tangents and adjoints of the implemented
       adaptive residual;
 \item a Rankine--Hugoniot-constrained graph representation of the principal
       reflected front and an independent classification of secondary ridges;
 \item a qualified fixed-$a$ potential-flow sensitivity
       $H_{2,a}^{\PF}(0.5;1.4)=-0.217\pm0.012$;
 \item an exact geometric back-map separating the canonical ratio from the
       physical trajectory angle and an exact chain rule from fixed $a$ to
       fixed $\lambda=(M-1)/\alpha^2$;
 \item a matched-boundary full-Euler study in which
       $\rho=h_\eta/\sqrt\mu$ is refined together with $\mu$;
 \item numerical evidence that the leading Euler and UTSD limits are not
       distinguishable within the qualified numerical envelope; and
 \item a direct same-strength grid-phase audit showing why the Euler cubic
       coefficient cannot yet be extrapolated uniquely.
\end{enumerate}

To keep the main argument readable, the main text gives only the equations and
algorithmic choices needed to interpret the principal results.  Appendices
\ref{app:euler-reduction}--\ref{app:reproducibility} provide the direct
Euler--UTSD reduction, complete coordinate and parameter-path derivations,
finite-strength residual, shock variations, potential-flow and Euler
finite-volume schemes, characteristic boundaries, continuation and remapping,
front extraction, topology tests, asymptotic-preserving design, phase audit,
and numerical uncertainty hierarchy.  The appendices are intended to make
every displayed coefficient independently auditable without requiring the
source code.

\section{Canonical self-similar problem and trajectory map}
\label{sec:canonical}

With the potential-flow convention
\begin{equation}
 u=\Phi_x,\qquad v=\Phi_y,
 \qquad u_t+u u_x+v_y=0,
 \qquad u_y-v_x=0,
 \label{eq:utsd-time}
\end{equation}
and self-similar coordinates $(\xi,\eta)=(x/t,y/t)$, the canonical system is
\begin{subequations}
\label{eq:ss-utsd}
\begin{align}
 (u-\xi)u_\xi-\eta u_\eta+v_\eta&=0,
 \label{eq:ss-utsd-a}\\
 u_\eta-v_\xi&=0.
 \label{eq:ss-utsd-b}
\end{align}
\end{subequations}
Equivalently,
\begin{subequations}
\begin{align}
 \partial_\xi\left(\frac12u^2-\xi u\right)
 +\partial_\eta(v-\eta u)+2u&=0,\\
 \partial_\xi(-v)+\partial_\eta u&=0.
\end{align}
\end{subequations}
The wall condition is $v(\xi,0)=0$.  At $\mu=0$, the incident shock and the
uniform state immediately behind it are
\begin{equation}
 \xi=a\eta+\frac12+a^2,
 \qquad (u,v)=(1,-a),
 \label{eq:incident-canonical}
\end{equation}
with $(u,v)=(0,0)$ ahead.  The canonical sonic diagnostic is
\begin{equation}
 D[u]=u-\xi-\frac14\eta^2.
 \label{eq:sonic}
\end{equation}

Let $(\xi_T,\eta_T)$ denote the fitted attachment of the principal reflected
front to the incident front.  The canonical functional studied here is
\begin{equation}
 G_{\rm can}=\frac{\eta_T}{\xi_T}.
 \label{eq:g-functional}
\end{equation}
It is deliberately distinguished from the physical ray angle.  The exact map
between canonical and physical self-similar coordinates is
\begin{equation}
 X_T=1+\frac{\mu}{2}\xi_T,
 \qquad
 Y_T=\frac{\sqrt\mu}{2}\eta_T,
 \label{eq:coordinate-map}
\end{equation}
and hence
\begin{equation}
 \theta_{\rm phys}
 =\operatorname{atan2}\!\left(
 \frac{\sqrt\mu}{2}\eta_T,
 1+\frac{\mu}{2}\xi_T
 \right).
 \label{eq:physical-angle}
\end{equation}
If
$\xi_T=\xi_0+\mu\xi_2+\OO(\mu^2)$ and
$\eta_T=\eta_0+\mu\eta_2+\OO(\mu^2)$, then
\begin{equation}
 \theta_{\rm phys}
 =G_0\del+K_2\del^3+\OO(\del^5),
 \label{eq:physical-series}
\end{equation}
where
\begin{equation}
 G_0=\frac{\eta_0}{2},
 \qquad
 K_2=\frac{\eta_2}{2}
 -\frac{\eta_0\xi_0}{4}
 -\frac{\eta_0^3}{24}.
 \label{eq:physical-coefficients}
\end{equation}
Thus neither $g=\eta_T/\xi_T$ nor its derivative is, without
\cref{eq:coordinate-map}, a physical angle coefficient.

\subsection{Fixed-\texorpdfstring{$a$}{a} and fixed-\texorpdfstring{$\lambda$}{lambda} parameter paths}
\label{sec:parameter-path}

The parent calculation below differentiates at fixed
$a=\alpha/\sqrt\mu$.  The classical distinguished limit is often stated
instead at fixed
\begin{equation}
 \lambda=\frac{M-1}{\alpha^2}.
 \label{eq:lambda-definition}
\end{equation}
Since $M=\sqrt{1+\mu/2}$ and $\alpha=a\sqrt\mu$,
\begin{equation}
 \lambda(a,\mu)=
 \frac{1}{2a^2\left(\sqrt{1+\mu/2}+1\right)}.
 \label{eq:lambda-a-mu}
\end{equation}
Thus a fixed-$a$ path approaches
$\lambda_0=1/(4a^2)$ but does not hold $\lambda$ exactly constant.  Conversely,
for a prescribed $\lambda$,
\begin{equation}
 a(\mu;\lambda)=
 \left[2\lambda\left(\sqrt{1+\mu/2}+1\right)\right]^{-1/2}
 =a_0\left(1-\frac{\mu}{16}\right)+\OO(\mu^2),
 \qquad a_0=\frac{1}{2\sqrt\lambda}.
 \label{eq:a-fixed-lambda}
\end{equation}
The canonical and physical cubic coefficients on the two paths are therefore
related by
\begin{subequations}
\label{eq:path-chain-rule}
\begin{align}
 H_{2,\lambda}(a_0;\gamma)
 &=H_{2,a}(a_0;\gamma)
 -\frac{a_0}{16}g'(a_0),
 \label{eq:path-chain-rule-g}\\
 K_{2,\lambda}(a_0;\gamma)
 &=K_{2,a}(a_0;\gamma)
 -\frac{a_0}{16}G_0'(a_0).
 \label{eq:path-chain-rule-G}
\end{align}
\end{subequations}
Equations \eqref{eq:path-chain-rule-g}--\eqref{eq:path-chain-rule-G} are exact
at the displayed order.  The present study qualifies the fixed-$a$
derivatives.  A numerical fixed-$\lambda$ coefficient additionally requires a
separately converged incidence derivative of the leading branch and is not
reported here.

\section{Finite-strength potential-flow parent}
\label{sec:parent}

We start from nondimensional isentropic potential flow,
\begin{align}
 \rho_T+\nabla_{X,Y}\!\cdot(\rho\nabla\varphi)&=0,
 \label{eq:mass}\\
 \varphi_T+\frac12|\nabla\varphi|^2
 +\frac{\rho^{\gamma-1}-1}{\gamma-1}&=0.
 \label{eq:bernoulli}
\end{align}
Set $\eps=M^2-1$, so $\mu=2\eps$, and introduce
\begin{equation}
 x=\frac{X-T}{\eps},
 \qquad y=\frac{\sqrt2\,Y}{\sqrt\eps},
 \qquad t=T,
 \qquad
 \varphi=\frac{2\eps^2}{\gamma+1}\Phi(x,y,t).
 \label{eq:weak-scaling}
\end{equation}
A direct density expansion gives a normalised residual
\begin{equation}
 \mathcal R_\eps
 =\mathcal R_0+\eps\mathcal R_1+\OO(\eps^2),
 \qquad
 \mathcal R_0=u_t+u u_x+v_y,
 \label{eq:residual-expansion-eps}
\end{equation}
so that
\begin{equation}
 F_2=\left.\partial_\mu\mathcal R\right|_{\mu=0}
 =\frac12\mathcal R_1.
 \label{eq:F2}
\end{equation}
For $\Phi=t\phi(\xi,\eta)$ and
$P=\Phi_t=\phi-\xi u-\eta v$, the smooth correction can be written in
conservative form,
\begin{equation}
 F_2^{(1)}=\partial_\xi f_2+\partial_\eta g_2+s_2,
 \qquad F_2^{(2)}=0,
 \label{eq:F2-conservative}
\end{equation}
with the explicit fluxes and source recorded in \cref{app:parent}.  This form
is used directly in the shock-capturing discretisation.

At fixed $a$, the exact finite-strength incident shock is
\begin{equation}
 \xi=A(a,\mu)\eta+C(a,\mu),
 \label{eq:exact-incident}
\end{equation}
where
\begin{align}
 A(a,\mu)&=\frac{\tan(a\sqrt\mu)}{\sqrt\mu},\\
 C(a,\mu)&=\frac{2}{\mu}
 \left[
 \frac{\sqrt{1+\mu/2}}{\cos(a\sqrt\mu)}-1
 \right].
 \label{eq:exact-incident-AC}
\end{align}
The removable limits reproduce \cref{eq:incident-canonical}.  To first order,
\begin{align}
 A&=a+\frac{a^3}{3}\mu+\OO(\mu^2),\\
 C&=\frac12+a^2
 +\mu\left(-\frac1{16}+\frac{a^2}{4}+\frac{5a^4}{12}\right)
 +\OO(\mu^2),\\
 u_b&=1-\left(\frac14+\frac{a^2}{2}\right)\mu+\OO(\mu^2),\\
 v_b&=-a+\left(\frac a4+\frac{a^3}{6}\right)\mu+\OO(\mu^2).
 \label{eq:incident-expansions}
\end{align}
The implemented derivative therefore includes both the interior term
\cref{eq:F2-conservative} and the differentiated incident geometry and state.

\begin{assumption}[Regular parent branch]
At fixed $(a,\gamma)$, the phase-fixed residual, its boundary data, and the
chosen canonical functional are differentiable in $\mu$ near $\mu=0$, and the
linearised residual is invertible on the selected Guderley branch.
\end{assumption}
Under this assumption, the implicit-function theorem gives
\cref{eq:canonical-expansion}.  The odd-power structure in
\cref{eq:physical-series} follows only after the exact physical back-map is
applied.

\section{Free-boundary linearisation and discrete adjoint}
\label{sec:adjoint}

For a perturbation $(w,z)$ of a smooth canonical state $(u_0,v_0)$, the
linearised operator is
\begin{equation}
 \mathcal L_0
 \begin{pmatrix}w\\z\end{pmatrix}
 =
 \begin{pmatrix}
 (u_0-\xi)w_\xi+(u_{0,\xi}-\eta\partial_\eta)w+z_\eta\\
 w_\eta-z_\xi
 \end{pmatrix}.
 \label{eq:linear-operator}
\end{equation}
For smooth test functions $(p,q)$, integration by parts gives the formal
adjoint
\begin{equation}
 \mathcal L_0^*
 \begin{pmatrix}p\\q\end{pmatrix}
 =
 \begin{pmatrix}
 -(u_0-\xi)p_\xi+\eta p_\eta+2p-q_\eta\\
 -p_\eta+q_\xi
 \end{pmatrix}
 \label{eq:continuous-adjoint}
\end{equation}
and the boundary form
\begin{equation}
 \mathscr B=
 w\{p(u_0-\xi)n_\xi+(-\eta p+q)n_\eta\}
 +z(pn_\eta-qn_\xi).
 \label{eq:boundary-form}
\end{equation}
On a fitted shock, the two traces of \cref{eq:boundary-form} combine with the
linearised Rankine--Hugoniot and shape equations.  In the discrete method all
such wall, boundary-motion, and shock terms are included in the transpose of
the assembled residual rather than inserted separately.

Let the complete phase-fixed residual and objective be
\begin{equation}
 \mathbb R(\bU,a,\mu)=0,
 \qquad G=G(\bU,a,\mu).
\end{equation}
The objective adjoint is
\begin{equation}
 \mathbb J^*\bPsi=G_{\bU}^*,
 \label{eq:objective-adjoint}
\end{equation}
which gives
\begin{equation}
 H_{2,a}=G_\mu-\ip{\bPsi}{\mathbb R_\mu}.
 \label{eq:H2-continuous}
\end{equation}
The discrete counterpart is exact for the implemented residual.  At the
converged state $\bx_0$,
\begin{equation}
 J_h=\frac{\partial\bR_h}{\partial\bx},
 \qquad
 J_h^T\bm\psi_h=G_{\bx,h}^T,
 \qquad
 H_{2,a,h}=G_{\mu,h}-\bm\psi_h^T\bR_{\mu,h}.
 \label{eq:discrete-adjoint}
\end{equation}
Jacobian--vector and transpose-Jacobian--vector products are obtained by
algorithmic differentiation.  The tangent and adjoint therefore differentiate
the same boundary construction, AMR refluxing, and trajectory objective that
enter the primal solve.

\section{Shock-fitted adaptive potential-flow calculation}
\label{sec:pf-method}

The canonical and parent problems are discretised by a second-order
finite-volume method on a three-level, conservatively refluxed nested hierarchy
\citep{BergerColella1989}.  Pseudo-transient Newton--Krylov iteration is
preconditioned by a multilevel sparse ILU approximation.  The principal
reflected front is promoted to a low-dimensional graph.  Writing the exact
incident front as $\xi_I=A\eta+C$, the lower branch is represented by
\begin{equation}
 \xi_s(\eta)=\xi_I(\eta)
 +\sum_{k=1}^p c_k z^k,
 \qquad
 z=\frac12\left(q+\sqrt{q^2+\epsilon_h^2}\right),
 \qquad q=\eta_T-\eta.
 \label{eq:fitted-graph}
\end{equation}
The smooth hinge makes the fitted branch coincide with the incident shock
above the attachment.

For the conservative UTSD fluxes
\begin{equation}
 F(U;\xi)=
 \begin{pmatrix}\frac12u^2-\xi u\\-v\end{pmatrix},
 \qquad
 G(U;\eta)=
 \begin{pmatrix}v-\eta u\\u\end{pmatrix},
\end{equation}
the graph jump law is
\begin{equation}
 \jump{F}-\xi_s'(\eta)\jump{G}=0.
 \label{eq:graph-RH}
\end{equation}
States are sampled at fixed canonical normal offsets, and the graph parameters
minimise
\begin{equation}
 \mathcal J(\theta;\bx,\mu)=\frac12\left(
 w_d^2\|r_{\rm ridge}\|^2
 +w_{RH}^2\|r_{RH}\|^2
 +w_c^2\|r_{\rm curvature}\|^2
 +w_I^2\|r_{\rm incident}\|^2
 \right).
 \label{eq:fit-cost}
\end{equation}
The stationarity equations $K=\partial_\theta\mathcal J=0$ are differentiated
together with the global residual.  Twelve admissible combinations of normal
offset and data/RH weight are retained as a functional family; their spread is
treated as a methodological, not statistical, uncertainty.

\begin{figure}[t]
 \centering
 \includegraphics[width=0.82\linewidth]{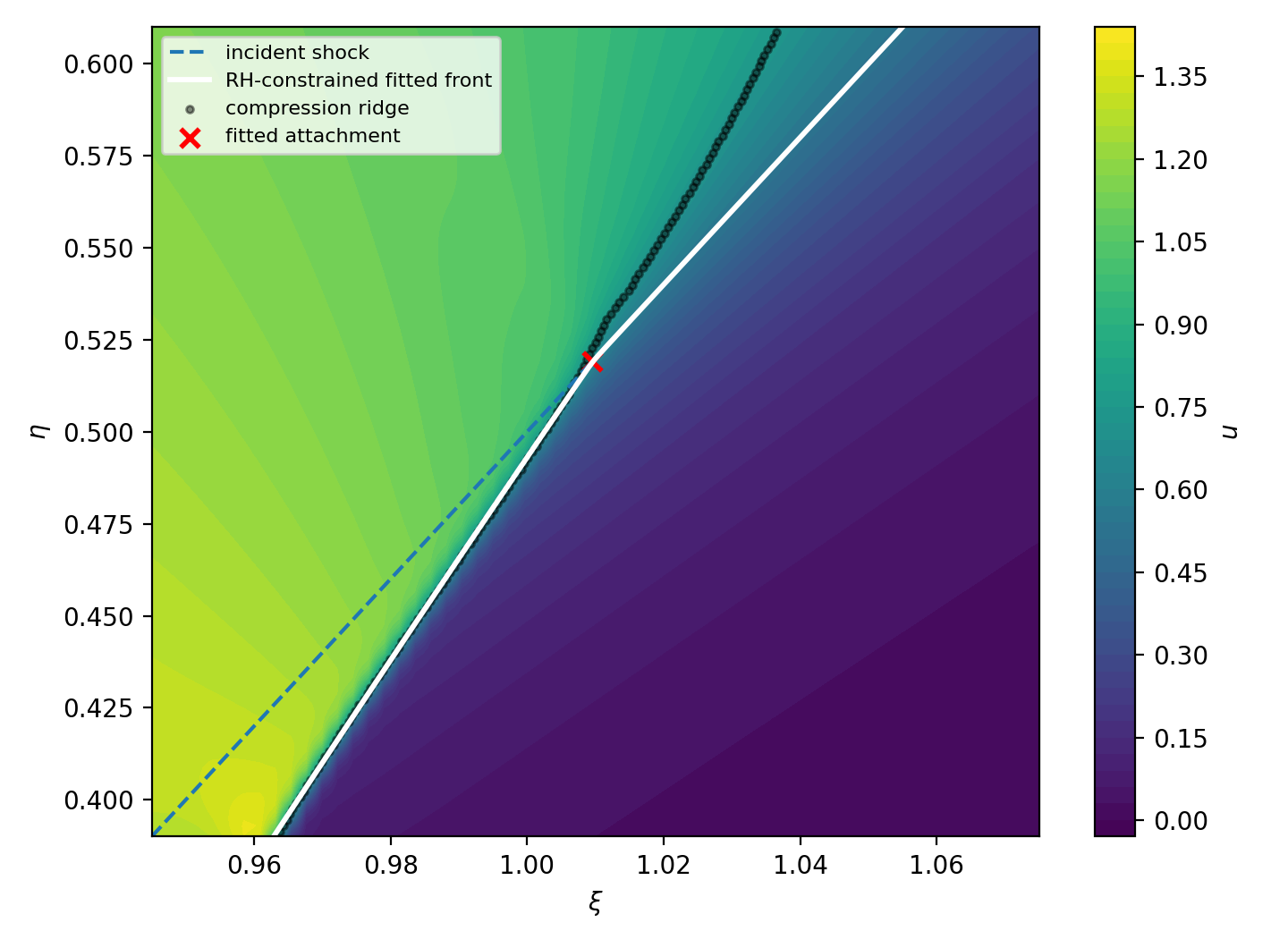}
 \caption{Effective-ratio-48 potential-flow field, exact incident shock,
 compression ridge, and Rankine--Hugoniot-constrained fitted principal front.
 The marker denotes the fitted attachment used in the canonical trajectory
 functional.}
 \label{fig:pf-front}
\end{figure}

Candidate secondary ridges are extracted in incident-shock coordinates,
clustered into continuous branches, and tested by the normalised defect of
\cref{eq:graph-RH}.  Two secondary compression fronts pass the jump-law audit;
two additional ridges are fan-like and are not forced into shock fits.  This
classification prevents a nearby secondary ridge from being mistaken for the
leading attachment.

\section{Potential-flow verification and finite-strength sensitivity}
\label{sec:pf-results}

The density expansion, conservative parent residual, and incident-shock
relations were checked symbolically.  Independent array and differentiable
implementations agree to roundoff on manufactured states.  The refluxed
Jacobian/transpose dot-product identity is satisfied at relative errors of
order $10^{-15}$.  At effective ratio 40, the exact fitted-front tangent and
complete finite-strength re-solves differ by approximately $0.14\%$.

\begin{table}[t]
\centering
\caption{Shock-fitted potential-flow results at $a=0.5$ and $\gamma=1.4$.
Standard deviations are over the 12-member fitted-front functional family.}
\label{tab:pf-results}
\begin{tabular}{rrrrr}
\toprule
Effective ratio & mean $g_h$ & SD$(g_h)$ & mean $H_{2,a,h}$ & SD$(H_{2,a,h})$\\
\midrule
32 & 0.515638 & 0.000573 & -0.217099 & 0.007603\\
40 & 0.514874 & 0.000328 & -0.213414 & 0.007190\\
48 & 0.514308 & 0.000194 & -0.221445 & 0.003322\\
\bottomrule
\end{tabular}
\end{table}

The leading values decrease smoothly under refinement, while the three finest
correction estimates cluster near $-0.217$.  Combining between-grid variation
with the largest within-grid functional spread gives
\begin{equation}
 \boxed{H_{2,a}^{\PF}(0.5;1.4)=-0.217\pm0.012.}
 \label{eq:H2a-pf}
\end{equation}
The interval is a conservative numerical-method envelope.  It is not a
confidence interval and does not include a model-form error associated with
replacing full Euler by isentropic potential flow.

\begin{figure}[t]
 \centering
 \includegraphics[width=0.74\linewidth]{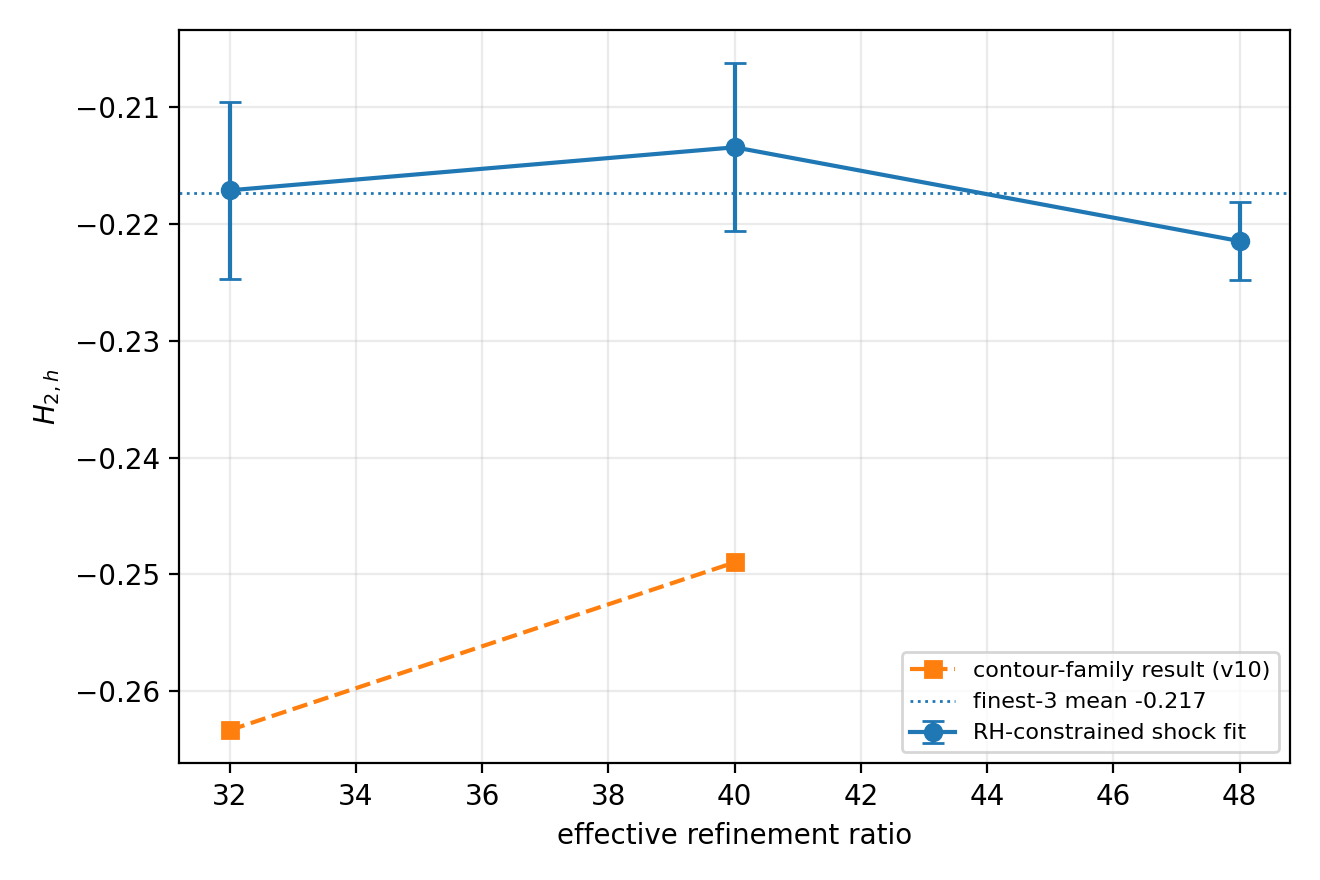}
 \caption{Potential-flow finite-strength sensitivity from the
 Rankine--Hugoniot-constrained fitted front.  Error bars show the spread over
 admissible fit settings.  The earlier contour-family sequence is included to
 show the trajectory-definition drift removed by fitting the discontinuity.}
 \label{fig:H2-convergence}
\end{figure}

The physical cubic coefficient is obtained from the separate derivatives of
$\xi_T$ and $\eta_T$ through \cref{eq:physical-coefficients}.  At effective
ratio 40 this gives
\begin{equation}
 K_{2,a}^{\PF}\simeq-0.266.
 \label{eq:K2a-pf}
\end{equation}
Because only one fully differentiated physical back-map is presently used,
\cref{eq:K2a-pf} is quoted as a diagnostic rather than with an independent
continuum uncertainty.

\section{Full-Euler correspondence and asymptotic-preserving limit}
\label{sec:euler}

\subsection{Continuous Euler--UTSD map}

The self-similar Euler equations can be written as
\begin{equation}
 \partial_X\{\bm F(\bm Q)-X\bm Q\}
 +\partial_Y\{\bm G(\bm Q)-Y\bm Q\}
 +2\bm Q=0.
 \label{eq:ss-euler}
\end{equation}
The coordinate and velocity map is
\begin{equation}
 X=1+\frac\mu2\xi,
 \qquad Y=\frac{\sqrt\mu}{2}\eta,
 \qquad
 U=\frac{\mu}{\gamma+1}u,
 \qquad V=\frac{\mu^{3/2}}{\gamma+1}v.
 \label{eq:euler-map}
\end{equation}
Direct expansion of the mass and longitudinal momentum equations gives
\begin{equation}
 (u-\xi)u_\xi-\eta u_\eta+v_\eta=0,
 \qquad
 u_\eta-v_\xi=0
 \label{eq:euler-to-utsd}
\end{equation}
at the first non-trivial order.  The exact incident geometry
\cref{eq:exact-incident} and post-shock state reduce to
\cref{eq:incident-canonical}, and the Euler pseudo-sonic discriminant reduces
to \cref{eq:sonic}.  The continuous equations therefore contain no order-one
Euler--UTSD scaling mismatch.

\subsection{Boundary problem and numerical method}

A subsonic outer boundary must not fix all conservative variables.  The Euler
calculation imposes only incoming characteristic information, using the
linearised reflected-wave outer solution on the subsonic left and top
boundaries; outgoing characteristics are retained from the interior.  The
wall is slip, the incident shock is imposed through its exact finite-strength
geometry, and the top boundary switches to the incident solution in the
supersonic region.

The Euler residual uses a dimensionally split, second-order MUSCL ALE--HLL
finite-volume scheme \citep{HartenLaxVanLeer1983}.  Wave speeds are measured
relative to the self-similar face velocity.  Strength continuation and
conservative interpolation connect neighbouring $\mu$ values and grids.  The
nonlinear corrector uses a sparse exact Jacobian assembled by stencil colouring
and a regularised Newton solve.  Every retained state has RMS residual at or
below $2\times10^{-6}$.

The relevant stretched-resolution parameter is
\begin{equation}
 \rho=\frac{h_\eta}{\sqrt\mu}.
 \label{eq:rho}
\end{equation}
A fixed-grid limit makes $\rho$ grow as $\mu\to0$ and therefore resolves the
transverse weak-shock structure progressively less well.  The qualified study
instead follows approximately constant-$\rho$ sequences and then extrapolates
$\rho\to0$.

\subsection{Trajectory family, topology, and coupled models}

All Euler sequences in this paper also hold $a=0.5$ fixed.  This is the
correct path for comparison with the fixed-$a$ potential-flow derivative and
for determining the common leading limit; it is not an exact fixed-$\lambda$
sequence at cubic order.  The Euler front is evaluated with the same 12
combinations of fixed canonical
normal offsets and Rankine--Hugoniot weights at every state.  Both
$(\xi_T,\eta_T)$ and the exact angle \cref{eq:physical-angle} are stored.  A
state is admitted only if the principal compression ridge retains the same
connectivity to the incident front, wall, and sonic set.  Across the final
dataset, the audit detects one continuous principal branch and no branch
switch.

The final set comprises 21 states and 252 front-functional evaluations:
\begin{table}[t]
\centering
\caption{Qualified Euler resolution slices.  The finest slice contains the
three weakest strengths; the other slices contain six strengths each.}
\label{tab:euler-slices}
\begin{tabular}{lrrrr}
\toprule
Slice & states & $\rho_{\min}$ & $\rho_{\max}$ & $\mu$ range\\
\midrule
coarse AP & 6 & 0.1172 & 0.1198 & 0.025--0.100\\
medium AP & 6 & 0.0978 & 0.1002 & 0.025--0.100\\
fine AP & 6 & 0.0721 & 0.0750 & 0.025--0.100\\
finest AP & 3 & 0.0586 & 0.0641 & 0.025--0.040\\
\bottomrule
\end{tabular}
\end{table}

For a response $Y\in\{g,\theta_{\rm phys}/\sqrt\mu,\xi_T,\eta_T\}$, the most
general retained coupled model is
\begin{equation}
 Y(\mu,\rho)=Y_0+a_1\mu+a_2\mu^2+b_1\rho
 +b_2\rho\sqrt\mu+b_3\rho^2,
 \label{eq:joint-model}
\end{equation}
together with restricted submodels.  Models are compared using corrected
Akaike information criteria, pointwise leave-one-out errors,
leave-one-resolution-slice-out errors, restriction to $\mu\leq0.05$, and all
12 trajectory definitions.  The quoted uncertainty envelopes include the
spread over model form, strength range, and front functional.

\subsection{Same-strength grid-phase audit}
\label{sec:phase-audit}

A captured shock can move by a fraction of a cell while the continuum front is
unchanged.  To measure this error directly, rather than infer it from
solutions at different strengths, we compute three neighbouring grids at each
of $\mu=0.05$ and $0.025$ for each of three stretched-resolution levels.  The
18 converged states are analysed with the identical 12-member functional
family, giving 216 front fits.  Within each bracket, the mean over the three
grids is used as a phase-averaged observable and the half-range is propagated
as a conservative phase envelope.  This audit is separate from the 21-state
leading-limit dataset and is designed specifically to test the cubic secant.

\section{Euler--UTSD comparison}
\label{sec:comparison}

\begin{figure}[t]
 \centering
 \includegraphics[width=0.84\linewidth]{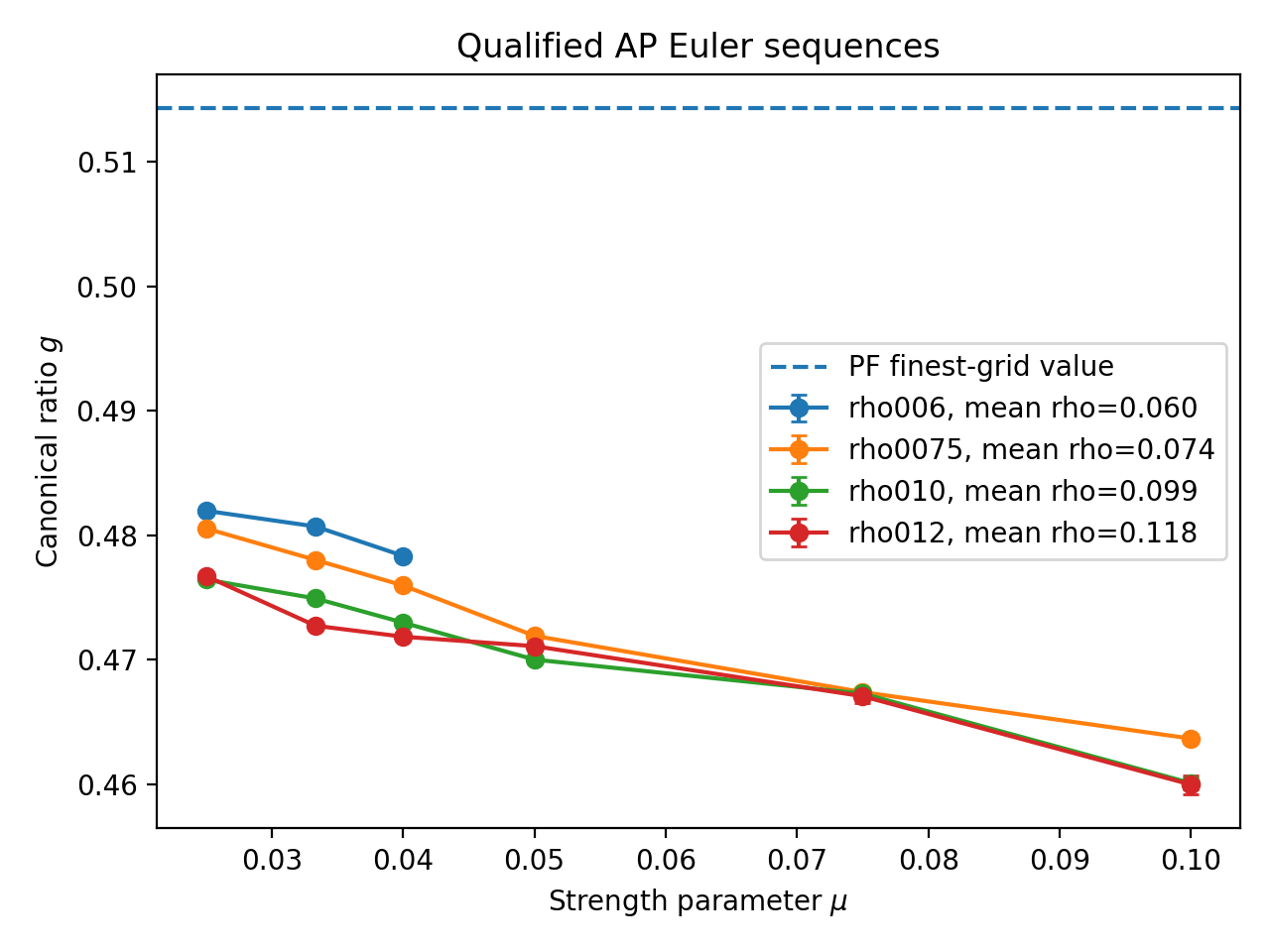}
 \caption{Canonical fitted-front ratio along the four
 strength-dependent-resolution Euler sequences.  The horizontal reference is
 the finest shock-fitted potential-flow value.}
 \label{fig:euler-canonical}
\end{figure}

The coupled extrapolation gives
\begin{equation}
 \boxed{g_0^{\Eul}=0.510\pm0.006,}
 \label{eq:g-euler}
\end{equation}
where the envelope is numerical and methodological.  The shock-fitted
potential-flow sequence is
\begin{equation}
 g_h^{\PF}=0.515638,\ 0.514874,\ 0.514308
 \label{eq:g-pf-sequence}
\end{equation}
at effective ratios 32, 40, and 48.  Fits with common assumed orders
$p=1/2,1,2$ span
\begin{equation}
 0.508\lesssim g_0^{\PF}\lesssim0.513,
 \label{eq:g-pf-envelope}
\end{equation}
with finest-grid value $0.5143$.  The Euler and potential-flow canonical
limits therefore overlap.

Applying the exact physical map gives
\begin{equation}
 \boxed{G_0^{\Eul}=0.256\pm0.004,}
 \qquad
 G_0^{\Eul}=\lim_{\mu\to0}\frac{\theta_{\rm phys}}{\sqrt\mu}.
 \label{eq:G-euler}
\end{equation}
The potential-flow sequence decreases from $0.26021$ to $0.25955$, with
common-order extrapolations
\begin{equation}
 0.257\lesssim G_0^{\PF}\lesssim0.259.
 \label{eq:G-pf-envelope}
\end{equation}
Thus no leading Euler--UTSD discrepancy is resolved within the qualified
envelopes.

\begin{figure}[t]
 \centering
 \includegraphics[width=0.84\linewidth]{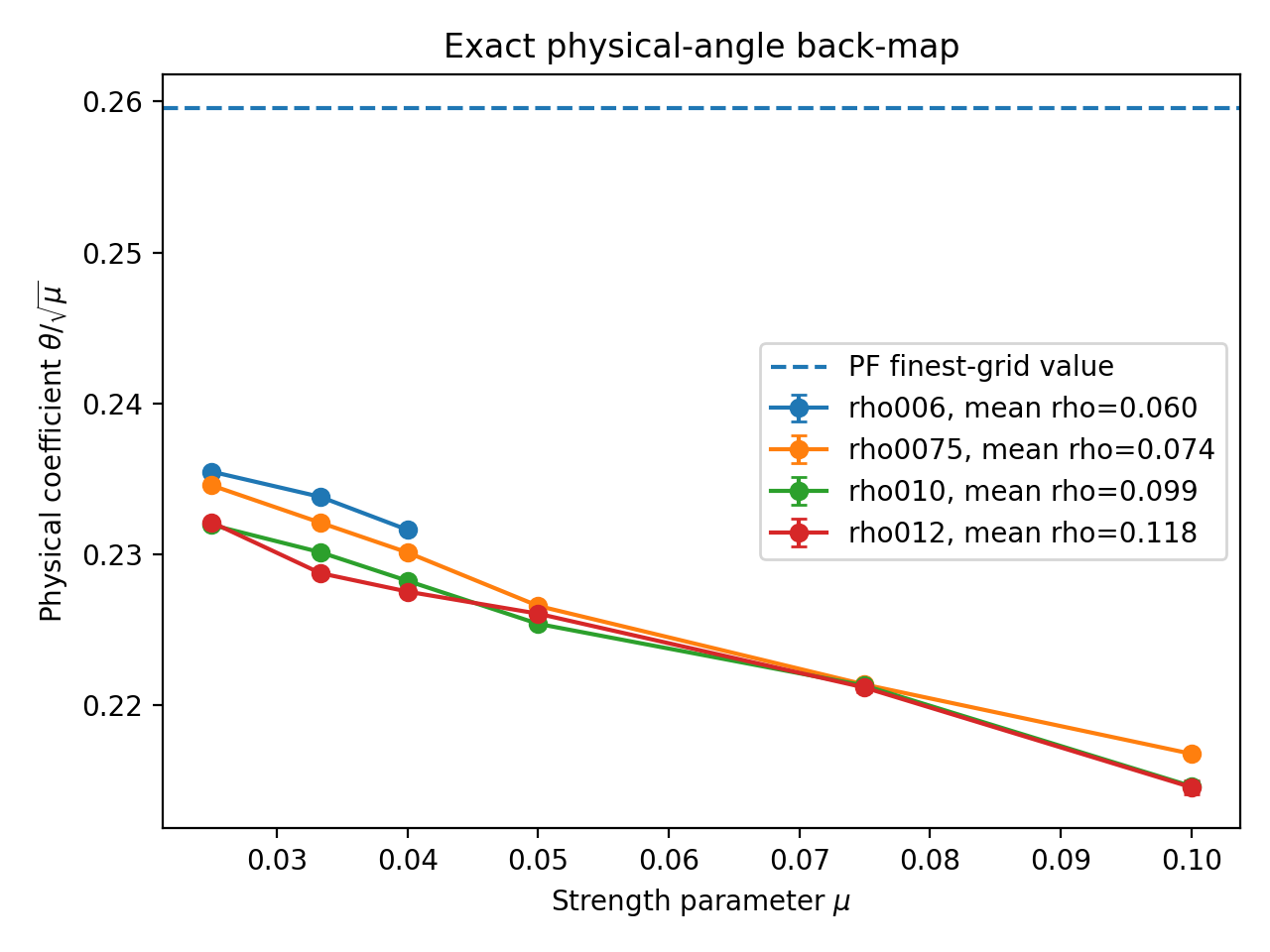}
 \caption{Physical trajectory coefficient obtained from the exact coordinate
 back-map.  The canonical ratio in \cref{fig:euler-canonical} is not used as a
 surrogate physical angle.}
 \label{fig:euler-physical}
\end{figure}

\begin{figure}[t]
 \centering
 \includegraphics[width=0.80\linewidth]{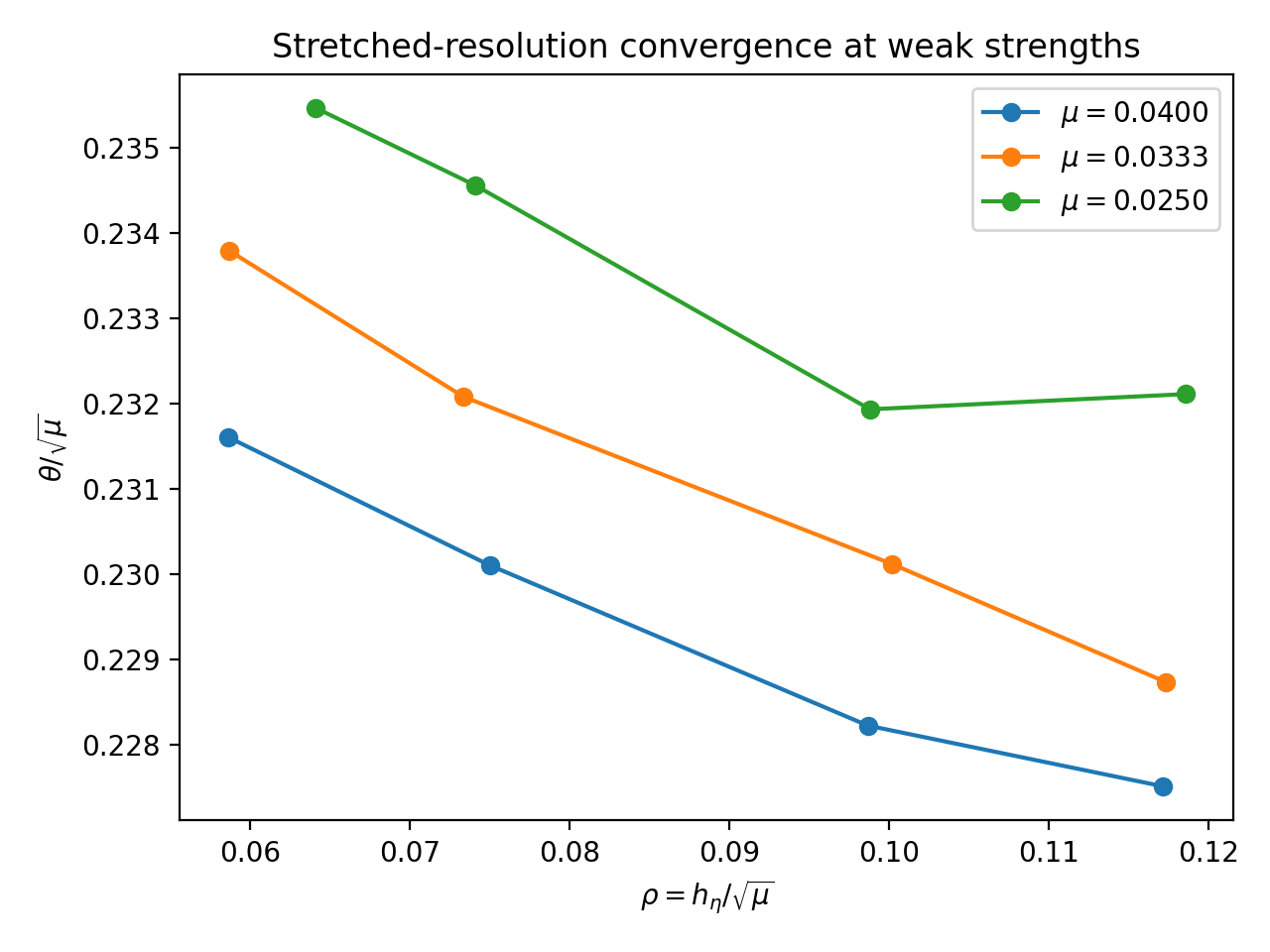}
 \caption{Dependence of the exact physical coefficient on the stretched mesh
 parameter $\rho=h_\eta/\sqrt\mu$ at the three weakest strengths shared across
 the resolution slices.}
 \label{fig:rho-convergence}
\end{figure}

The previously observed low leading intercept arose from taking $\mu\to0$ at
fixed $h_\eta$, so $\rho$ worsened, and from mixing
$\eta_T/\xi_T$ with the physical angle.  A further preliminary value near
$g=0.446$ at $\mu=0.05$ cannot be reproduced from the retained source plus the
documented matched characteristic boundary construction.  The reconstructed
matched branch instead plateaus near $g(0.05)=0.4711$.  The unreproduced value
is therefore excluded from the present analysis.

\subsection{Status of the Euler cubic coefficient}
\label{sec:euler-cubic}

The leading agreement does not qualify the Euler derivative with respect to
$\mu$.  The same-strength audit defines the physical secant
\begin{equation}
 K_{2,\mathrm{sec}}^{\Eul}(\rho)=
 \frac{G^{\Eul}(0.05,\rho)-G^{\Eul}(0.025,\rho)}{0.025}.
 \label{eq:euler-cubic-secant}
\end{equation}
The phase-bracketed values are
\begin{table}[t]
\centering
\caption{Same-strength phase-bracketed Euler cubic secants.  The uncertainty
is the conservative envelope obtained from the two three-grid half-ranges.}
\label{tab:phase-secants}
\begin{tabular}{rrrr}
\toprule
Mean $\rho$ & canonical secant & physical secant & physical phase envelope\\
\midrule
0.1193 & -0.2388 & -0.2507 & $\pm0.0282$\\
0.0988 & -0.2620 & -0.2657 & $\pm0.0395$\\
0.0735 & -0.3006 & -0.2918 & $\pm0.0354$\\
\bottomrule
\end{tabular}
\end{table}
Each finite-$\rho$ physical secant is compatible with the fixed-$a$
potential-flow diagnostic $K_{2,a}^{\PF}\simeq-0.266$.  However, the central
values drift systematically as $\rho$ decreases and do not form a resolution
plateau.

\begin{figure}[t]
 \centering
 \includegraphics[width=0.80\linewidth]{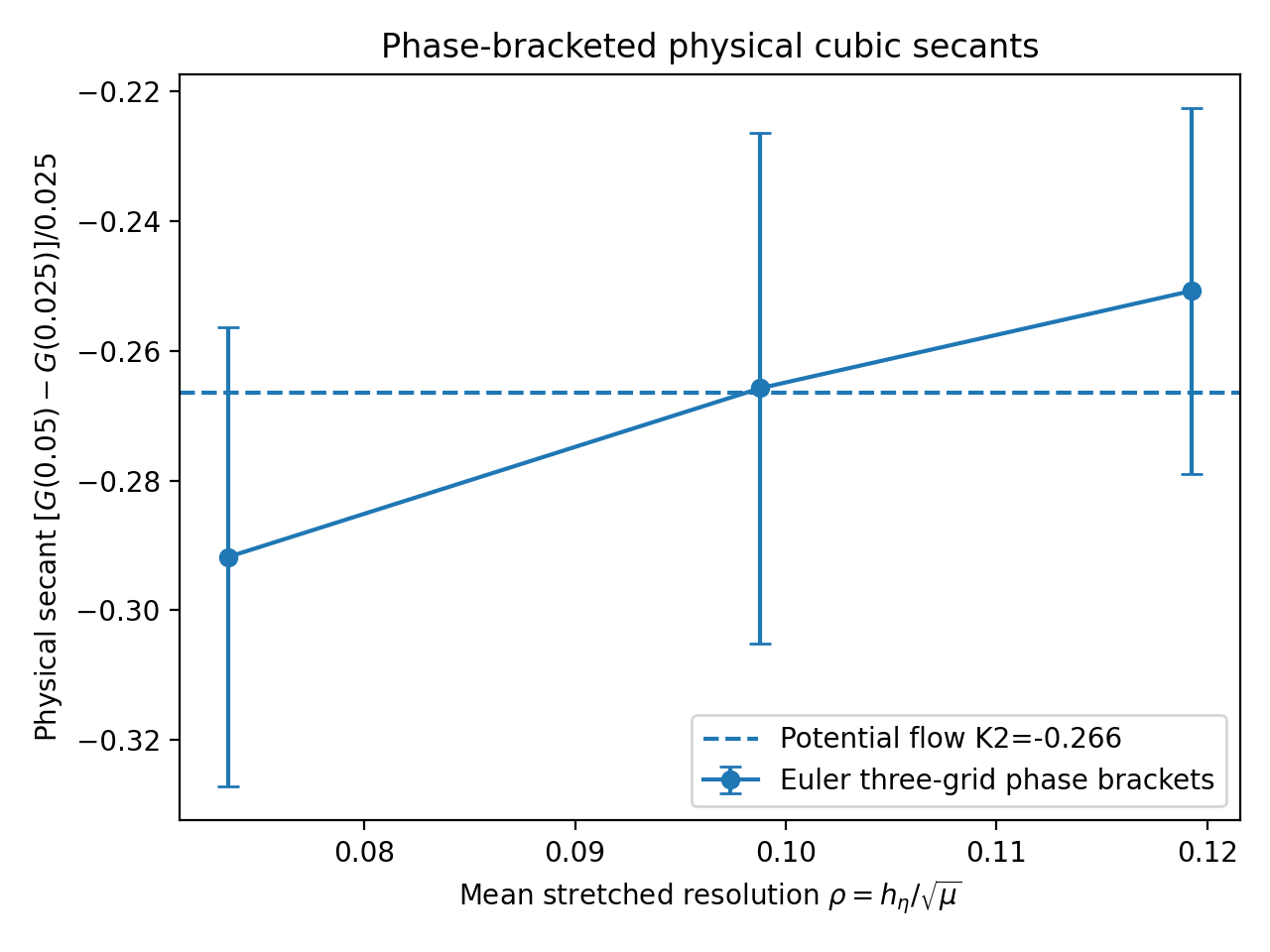}
 \caption{Same-strength, three-grid phase-bracketed physical Euler secants.
 The dashed line is the fixed-$a$ potential-flow diagnostic.  Individual
 finite-resolution values are compatible with potential flow, but the
 $\rho\to0$ trend is not converged.}
 \label{fig:phase-secants}
\end{figure}

A constant, linear, or quadratic extrapolation of the three central values
gives nominal physical limits $-0.2665$, $-0.3566$, and $-0.4149$,
respectively.  The constant model ignores the visible drift, the linear model
extrapolates beyond the sampled interval, and the quadratic model has no
predictive redundancy.  Explicit harmonic fits of the subcell phase likewise
return incompatible continuum secants.  Variations among the 12 smooth front
functionals are smaller than the grid-phase envelopes, so the obstruction is
not primarily the choice of trajectory functional.

Consequently,
\begin{equation}
 K_{2,a}^{\Eul}\ \text{is consistent with}\ K_{2,a}^{\PF},
 \qquad
 \text{but equality is not numerically verified.}
 \label{eq:cubic-verdict}
\end{equation}
Moreover, neither coefficient in this comparison is yet the fixed-$\lambda$
coefficient of \cref{eq:path-chain-rule-G}.  A definitive comparison requires
either a shock-fitted or front-tracked Euler formulation, or phase brackets at
substantially smaller $\rho$ with enough independent levels to identify the
continuum law.

\section{Discussion}
\label{sec:discussion}

The results separate four issues that had previously been conflated.

\paragraph{The leading canonical branch.}
The strength-dependent Euler sequences approach the shock-fitted UTSD branch
within the numerical envelope.  The topology audit finds no evidence that the
Euler computation follows a different leading compression ridge or a distinct
self-similar branch.

\paragraph{The role of asymptotic-preserving refinement.}
For this anisotropic weak-shock limit, $h_\eta$ and $\mu$ cannot be sent to zero
independently.  The fixed-grid sequence allowed $h_\eta/\sqrt\mu$ to grow and
therefore could converge to a discretisation-controlled limit.  The present
study does not prove that the split ALE--HLL scheme is uniformly
asymptotic-preserving for arbitrary data, but it controls the singular mesh
parameter directly and demonstrates convergence of the selected trajectory
functional.

\paragraph{Canonical versus physical observables.}
The canonical ratio $\eta_T/\xi_T$ is a useful and differentiable numerical
functional.  It is not the physical ray angle.  The exact back-map changes both
the leading and cubic coefficients.  Any comparison with experiment or a
finite-strength Euler calculation must therefore store the attachment
coordinates, not only their ratio.

\paragraph{Meaning of the potential-flow correction.}
Equation \eqref{eq:H2a-pf} is a qualified fixed-$a$ sensitivity of the adopted
isentropic potential-flow parent model.  It has passed primal, tangent,
adjoint, re-solve, grid, and front-functional tests.  Full-Euler leading
agreement supports the parent model at leading order, but the Euler cubic term
has not yet passed strength-range and model-form stability.  The potential-flow coefficient should therefore not be presented as a fully
Euler-validated physical constant.  It also should not be called the
fixed-$\lambda$ cubic coefficient until the chain correction in
\cref{eq:path-chain-rule} is evaluated.

The main remaining task is not another unstructured regression over weaker
captured shocks.  It is a shock-fitted or front-tracked Euler calculation, or a
substantially finer redundant phase-bracket hierarchy, carried along an exact
fixed-$\lambda$ path.  A separate converged calculation of $g'(a)$ and
$G_0'(a)$ is also required to convert the present parent sensitivity through
\cref{eq:path-chain-rule}.

\section{Conclusions}
\label{sec:conclusions}

We have derived and differentiated a finite-strength potential-flow parent for
the canonical Guderley--Mach reflection problem and tested its leading limit
against a matched-boundary full-Euler calculation.  The principal conclusions
are:
\begin{enumerate}[label=(\roman*)]
 \item the complete potential-flow parent has
       $R=R_0+\mu F_2+\OO(\mu^2)$, including interior, incident-front, and
       post-shock-state corrections;
 \item a Rankine--Hugoniot-constrained fitted front gives the qualified
       fixed-$a$ parent sensitivity
       $H_{2,a}^{\PF}(0.5;1.4)=-0.217\pm0.012$;
 \item the canonical and physical trajectory coefficients are distinct and
       are related by the exact map \cref{eq:physical-angle};
 \item the fixed-$a$ and fixed-$\lambda$ cubic coefficients are distinct and
       obey the exact chain rule \cref{eq:path-chain-rule};
 \item a 21-state, four-resolution Euler study with controlled
       $h_\eta/\sqrt\mu$ gives
       $g_0^{\Eul}=0.510\pm0.006$ and
       $G_0^{\Eul}=0.256\pm0.004$;
 \item these values overlap the shock-fitted potential-flow limits, so no
       leading Euler--UTSD discrepancy is resolved; and
 \item an independent 18-state, same-strength phase audit finds Euler cubic
       secants compatible with the potential-flow correction at finite
       resolution, but no stable $\rho\to0$ coefficient.
\end{enumerate}
The publication-level result is therefore leading-order Euler--UTSD
correspondence and a qualified fixed-$a$ potential-flow sensitivity.  Neither
a converged Euler cubic coefficient nor a fixed-$\lambda$ cubic coefficient is
claimed.

\section*{Data and software availability}
The supplementary research archive accompanying this manuscript contains the
source code, nonlinear checkpoints, trajectory-family data, model-comparison
tables, and figure-generation scripts.  The compact manuscript archive contains
the CSV and JSON files required to
reproduce the leading-limit tables, the same-strength phase brackets, and all
displayed extrapolation-sensitivity results.

\section*{Competing interests}
The author declares no competing interests.

\appendix

\section{Direct reduction of self-similar Euler to UTSD}
\label{app:euler-reduction}

This appendix fixes the normalisation used throughout the paper and records a
direct leading-order reduction from the compressible Euler equations.  Let
\begin{equation}
 \bm Q=(\rho,\rho U,\rho V,E)^T,
 \qquad
 E=\frac{p}{\gamma-1}+\frac12\rho(U^2+V^2),
\end{equation}
and write the nondimensional Euler equations as
\begin{equation}
 \bm Q_T+\partial_{\widehat X}\bm F(\bm Q)
 +\partial_{\widehat Y}\bm G(\bm Q)=0,
 \label{eq:app-euler-conservation}
\end{equation}
where the upstream density and sound speed are one, so that the upstream
pressure is $1/\gamma$.  For a self-similar state
$\bm Q(\widehat X,\widehat Y,T)=\bm Q(X,Y)$ with
$(X,Y)=(\widehat X/T,\widehat Y/T)$, the chain rule gives
\begin{equation}
 \partial_X\{\bm F(\bm Q)-X\bm Q\}
 +\partial_Y\{\bm G(\bm Q)-Y\bm Q\}+2\bm Q=0.
 \label{eq:app-selfsimilar-euler}
\end{equation}
This is an ALE conservation law whose grid velocity is $(X,Y)$.

Set $\mu=2(M^2-1)$ and introduce the canonical coordinates and velocities
\begin{equation}
 X=1+\frac{\mu}{2}\xi,
 \qquad
 Y=\frac{\sqrt\mu}{2}\eta,
 \qquad
 U=\frac{\mu}{\gamma+1}u,
 \qquad
 V=\frac{\mu^{3/2}}{\gamma+1}v.
 \label{eq:app-euler-scaling}
\end{equation}
On the weak, isentropic branch selected by the incident shock data,
\begin{equation}
 \rho=1+\frac{\mu}{\gamma+1}u+\OO(\mu^2),
 \qquad
 p=\frac1\gamma+\frac{\mu}{\gamma+1}u+\OO(\mu^2).
 \label{eq:app-thermo-expansion}
\end{equation}
Substitution of \cref{eq:app-euler-scaling,eq:app-thermo-expansion} into the
mass equation, division by the first non-zero common factor, and collection of
the leading terms gives
\begin{equation}
 (u-\xi)u_\xi-\eta u_\eta+v_\eta=0.
 \label{eq:app-leading-mass}
\end{equation}
The leading transverse-momentum compatibility is
\begin{equation}
 u_\eta-v_\xi=0.
 \label{eq:app-leading-curl}
\end{equation}
Equations \eqref{eq:app-leading-mass}--\eqref{eq:app-leading-curl} are exactly
\cref{eq:ss-utsd}.  Entropy production and vorticity generated by a weak shock
are higher-order effects on this branch and enter the finite-strength forcing,
not the leading UTSD equations.  The reduction also gives
\begin{equation}
 \frac{(U-X)^2+(V-Y)^2-c^2}{\mu/2}
 =-4\left(u-\xi-\frac{\eta^2}{4}\right)+\OO(\mu),
\end{equation}
so the Euler pseudo-sonic set reduces to \cref{eq:sonic}.  The exact incident
shock and post-shock state reduce to \cref{eq:incident-canonical}; hence the
continuous equations, incident data, and sonic diagnostic contain no hidden
order-one normalisation difference.

\section{Coordinate, observable, and parameter-path derivations}
\label{app:coordinate-derivation}

The physical self-similar location corresponding to a canonical attachment
$(\xi_T,\eta_T)$ is
\begin{equation}
 (X_T,Y_T)=\left(1+\frac\mu2\xi_T,\frac{\sqrt\mu}{2}\eta_T\right).
\end{equation}
The physical trajectory is the ray from the reflection point $(1,0)$ to the
attachment, not the ray from the canonical origin.  Therefore
\begin{equation}
 \theta_{\rm phys}
 =\arctan\left(\frac{Y_T}{X_T}\right)
 =\operatorname{atan2}\left(
 \frac{\sqrt\mu}{2}\eta_T,
 1+\frac\mu2\xi_T\right).
\end{equation}
Let
\begin{equation}
 \xi_T=\xi_0+\mu\xi_2+\OO(\mu^2),
 \qquad
 \eta_T=\eta_0+\mu\eta_2+\OO(\mu^2).
\end{equation}
Using $\arctan z=z-z^3/3+\OO(z^5)$ and expanding the denominator gives
\begin{align}
 \frac{Y_T}{X_T}
 &=\frac{\sqrt\mu}{2}\eta_0
 +\mu^{3/2}\left(\frac{\eta_2}{2}
 -\frac{\eta_0\xi_0}{4}\right)+\OO(\mu^{5/2}),\\
 \theta_{\rm phys}
 &=\frac{\eta_0}{2}\sqrt\mu
 +\left(\frac{\eta_2}{2}
 -\frac{\eta_0\xi_0}{4}-\frac{\eta_0^3}{24}\right)\mu^{3/2}
 +\OO(\mu^{5/2}).
\end{align}
This proves \cref{eq:physical-coefficients} and shows explicitly why
$\eta_T/\xi_T$ is not a physical angle coefficient.

The Euler calculations use
\begin{equation}
 M=\sqrt{1+\frac\mu2},
 \qquad
 \alpha=a\sqrt\mu.
\end{equation}
Consequently
\begin{equation}
 \lambda(a,\mu)=\frac{M-1}{\alpha^2}
 =\frac{1}{2a^2(\sqrt{1+\mu/2}+1)}.
\end{equation}
For exactly fixed $\lambda$, inversion gives
\begin{equation}
 a(\mu;\lambda)=
 [2\lambda(\sqrt{1+\mu/2}+1)]^{-1/2}
 =a_0\left(1-\frac\mu{16}+\OO(\mu^2)\right).
\end{equation}
If $G(a,\mu)=g(a)+\mu H_{2,a}(a)+\OO(\mu^2)$, then composition with
$a(\mu;\lambda)$ gives
\begin{equation}
 H_{2,\lambda}=H_{2,a}-\frac{a_0}{16}g'(a_0).
\end{equation}
The same argument applied to the physical coefficient gives
\cref{eq:path-chain-rule-G}.  The fixed-$\lambda$ correction is therefore a
leading-branch incidence derivative, not an error bar that can be inferred
from a fixed-$a$ calculation.

\section{Finite-strength potential-flow expansion}
\label{app:parent}

Starting from \cref{eq:mass,eq:bernoulli}, set $\eps=M^2-1=\mu/2$ and use
\cref{eq:weak-scaling}.  With $u=\Phi_x$, $v=\Phi_y$, and
\begin{equation}
 B=\varphi_T+\frac12|\nabla\varphi|^2,
 \qquad
 \rho=(1-(\gamma-1)B)^{1/(\gamma-1)},
\end{equation}
the Bernoulli quantity is
\begin{equation}
 B=-\frac{2\eps}{\gamma+1}u
 +\frac{2\eps^2}{\gamma+1}\Phi_t
 +\frac{2\eps^2}{(\gamma+1)^2}u^2
 +\frac{4\eps^3}{(\gamma+1)^2}v^2.
\end{equation}
Writing
$\rho=1+\eps r_1+\eps^2r_2+\eps^3r_3+\OO(\eps^4)$ gives
\begin{align}
 r_1&=\frac{2u}{\gamma+1},\\
 r_2&=-\frac{2}{(\gamma+1)^2}
 \left[(\gamma+1)P+(\gamma-1)u^2\right],\\
 r_3&=\frac{4}{3(\gamma+1)^3}
 \left[3(\gamma+1)(\gamma-2)Pu
 +2\gamma(\gamma-2)u^3-3(\gamma+1)v^2\right],
\end{align}
where $P=\phi-\xi u-\eta v$.  Expansion of mass conservation gives
$\mathcal R_\eps=\mathcal R_0+\eps\mathcal R_1+\OO(\eps^2)$ and hence
$F_2=\mathcal R_1/2$.  In conservative form,
\begin{equation}
 F_2^{(1)}=\partial_\xi f_2+\partial_\eta g_2+s_2,
 \qquad F_2^{(2)}=0,
\end{equation}
with
\begin{align}
 f_2&=\frac{\gamma+1}{8}
 \left(\frac{2r_2u}{\gamma+1}-r_3-\xi r_2\right),\\
 g_2&=\frac{\gamma+1}{8}
 \left(\frac{4r_1v}{\gamma+1}-\eta r_2\right),\\
 s_2&=\frac{\gamma+1}{4}r_2.
\end{align}
The potential $\phi$ required by these terms is reconstructed by integrating
both $u$ and $v$ and averaging the two path reconstructions; the mismatch is
retained as a compatibility diagnostic.

The exact finite-strength incident line follows from the planar shock with
normal $(\cos\alpha,-\sin\alpha)$ and speed $M$:
\begin{equation}
 \xi=A(a,\mu)\eta+C(a,\mu),
\end{equation}
where $A$ and $C$ are given in \cref{eq:exact-incident-AC}.  Taylor expansion
produces \cref{eq:incident-expansions}.  The derivative of the numerical
residual includes the moving incident graph, the changing boundary state, and
the interior forcing above; omitting any one of these terms differentiates a
different problem.

\section{Shock conditions, graph representation, and first variation}
\label{app:shock}

For the canonical conservation law
\begin{equation}
 \partial_\xi F(U;\xi)+\partial_\eta G(U;\eta)+S(U)=0,
\end{equation}
a stationary graph $\xi=S(\eta)$ satisfies
\begin{equation}
 [F]-S'[G]=0.
 \label{eq:app-graph-jump}
\end{equation}
For
\begin{equation}
 F=\begin{pmatrix}\frac12u^2-\xi u\\-v\end{pmatrix},
 \qquad
 G=\begin{pmatrix}v-\eta u\\u\end{pmatrix},
\end{equation}
this is equivalent to
\begin{subequations}
\begin{align}
 [v]+S'[u]&=0,\\
 \{u\}-S+(S')^2+\eta S'&=0.
\end{align}
\end{subequations}
Write
$S=S_0+\mu h$, $u=u_0+\mu w$, and $v=v_0+\mu z$.  Evaluation on the moving
graph gives the first variations
\begin{subequations}
\begin{align}
 [z]+h[v_{0,\xi}]+h'[u_0]
 +S_0'([w]+h[u_{0,\xi}])&=G_{2,1},\\
 \{w\}+h\{u_{0,\xi}\}-h+(2S_0'+\eta)h'&=G_{2,2},
\end{align}
\end{subequations}
where $G_{2,j}$ collect the explicit finite-strength flux and coordinate
forcing.  The fitted graph uses the smooth hinge
\begin{equation}
 z(\eta)=\frac12\left(q+\sqrt{q^2+\epsilon_h^2}\right),
 \qquad q=\eta_T-\eta,
\end{equation}
and
\begin{equation}
 S(\eta)=S_I(\eta)+\sum_{k=1}^{p}c_k z(\eta)^k.
\end{equation}
Above the attachment, $z$ is exponentially small relative to the local mesh
scale and the fitted branch joins the exact incident shock smoothly.

\section{Potential-flow finite-volume, AMR, and nonlinear solution}
\label{app:pf-numerics}

\subsection{Cell residual and numerical fluxes}

On a cell $C_{ij}$, the semidiscrete residual is
\begin{equation}
 R_{ij}=\frac{\widehat F_{i+1/2,j}-\widehat F_{i-1/2,j}}{\Delta\xi}
 +\frac{\widehat G_{i,j+1/2}-\widehat G_{i,j-1/2}}{\Delta\eta}
 +\begin{pmatrix}2u_{ij}\\0\end{pmatrix}
 +\mu R_{2,ij}.
\end{equation}
Piecewise-linear states are reconstructed using the three-argument minmod
limiter
\begin{equation}
 \sigma_i=\operatorname{minmod}(\vartheta\Delta_-U_i,
 \tfrac12(U_{i+1}-U_{i-1}),\vartheta\Delta_+U_i),
\end{equation}
with $\vartheta=1.25$ on the nested potential-flow hierarchy.  The leading
$x$-flux is
\begin{equation}
 \widehat F=\frac12(F_L+F_R)-\frac12a_x(U_R-U_L),
 \qquad
 a_x=\max\{1,|u_L-\xi_f|,|u_R-\xi_f|\},
\end{equation}
and the transverse flux uses the corresponding self-similar characteristic
bound
\begin{equation}
 a_\eta=\frac12\left(\eta_f+\sqrt{\eta_f^2+4}\right).
\end{equation}
The $\OO(\mu)$ physical correction uses arithmetic face fluxes while retaining
the leading dissipation.  This defines a differentiable truncated parent
residual whose added numerical-viscosity derivative vanishes under refinement.

\subsection{Nested conservative hierarchy}

The hierarchy consists of a uniform level 0, a rectangular level-1 patch
aligned with level-0 faces, and a level-2 patch aligned with level-1 faces.
The refinement ratios are four and two, respectively.  At a coarse--fine
interface, the coarse flux is replaced by the sum of the child fluxes over the
same physical face.  For covered parent cells, the PDE equation is replaced by
the conservative restriction constraint
\begin{equation}
 U^{(\ell)}_{ij}
 -\frac{1}{r^2}\sum_{m,n=0}^{r-1}
 U^{(\ell+1)}_{ri+m,rj+n}=0.
\end{equation}
Thus coarse and fine unknowns are solved simultaneously, refluxing is part of
the residual, and its tangent and transpose contain the same interface
couplings as the primal problem.

\subsection{Pseudo-transient Newton--Krylov solve}

Let $q$ contain all level unknowns.  At Newton iteration $k$, the regularised
step solves
\begin{equation}
 (J(q_k)+\sigma_k I)\Delta q_k=-R(q_k).
\end{equation}
Matrix-free Jacobian products are obtained by automatic differentiation,
GMRES is preconditioned with a multilevel sparse ILU approximation, and a
backtracking line search requires reduction of the residual norm.  The shift
$\sigma_k$ is reduced after successful Newton-like steps and increased when a
factorisation or line search fails.  The converged state must satisfy the PDE,
coarse--fine restriction, and reflux equations simultaneously.

\subsection{Fitted-front objective}

For each candidate graph, the residual vector contains a ridge-location block,
a normalised Rankine--Hugoniot block, a curvature regularisation, and an
incident-branch matching block.  The retained family uses fixed canonical
normal offsets
\begin{equation}
 d_n\in\{0.075,0.0875,0.1000,0.1125\}
\end{equation}
and RH weights
\begin{equation}
 w_{RH}\in\{0.75,1.0,1.5\},
\end{equation}
for 12 total functionals.  All grids use these same dimensional canonical
parameters; they are not specified in cell counts.

\section{Tangent, implicit front derivative, and discrete adjoint}
\label{app:adjoint-details}

Let the augmented unknown be $z=(q,\theta)$, where $q$ is the AMR state and
$\theta$ contains the attachment and graph coefficients.  Define
\begin{equation}
 \mathcal F(z,\mu)=
 \begin{pmatrix}R(q,\mu)\\K(\theta,q,\mu)\end{pmatrix}=0,
\end{equation}
where $K=\partial_\theta\mathcal J$ is the fitted-front stationarity system.
The exact tangent is
\begin{equation}
 \mathcal F_z z_\mu=-\mathcal F_\mu,
 \qquad
 \frac{\dd G}{\dd\mu}=G_\mu+G_z z_\mu.
\end{equation}
For the adjoint $\psi$,
\begin{equation}
 \mathcal F_z^T\psi=G_z^T,
 \qquad
 \frac{\dd G}{\dd\mu}=G_\mu-\psi^T\mathcal F_\mu.
 \label{eq:app-augmented-adjoint}
\end{equation}

The implementation eliminates the low-dimensional front fit first.  If
$r(\theta,q,\mu)$ is the fit residual, a Gauss--Newton approximation gives
\begin{equation}
 H_\theta=J_\theta^TJ_\theta,
 \qquad
 H_\theta^T\lambda=G_\theta^T,
 \qquad
 G_q^{\rm imp}=-(J_\theta^T\lambda)^T r_q.
\end{equation}
The resulting implicit objective gradient is the right-hand side of the global
transpose AMR solve.  Terms omitted by the Gauss--Newton reduction are
proportional to the fitted residual and vanish for an exact front fit; their
magnitude is monitored through the stationarity norm and Hessian condition
number.

The numerical verification hierarchy includes: (i) automatic-differentiation
versus independent residual implementations; (ii) Jacobian/transpose dot
products at relative error $\OO(10^{-15})$; (iii) centred differences of
$R_\mu$ and $G_\mu$; (iv) tangent--adjoint equality; and (v) complete
finite-strength re-solves.  At effective ratio 40 the fitted-front tangent and
re-solve derivative differ by approximately $0.14\%$.

\section{Full-Euler ALE--HLL discretisation and characteristic boundaries}
\label{app:euler-numerics}

\subsection{ALE flux and MUSCL--HLL residual}

For a face with outward unit normal $n$ and self-similar normal speed $w_n$,
the conservative ALE flux is
\begin{equation}
 \mathcal F_{\rm ALE}(Q;n,w_n)=\mathcal F(Q)\cdot n-w_n Q.
\end{equation}
For the vertical and horizontal faces, $w_n=X_f$ and $Y_f$, respectively.
Second-order one-dimensional MUSCL reconstructions use the minmod parameter
$\vartheta=1.1$.  Given left and right reconstructed states, define the
relative acoustic bounds
\begin{equation}
 S_L=\min(u_{n,L}-w_n-c_L,u_{n,R}-w_n-c_R),
 \qquad
 S_R=\max(u_{n,L}-w_n+c_L,u_{n,R}-w_n+c_R).
\end{equation}
The HLL flux is
\begin{equation}
 \widehat F_{\rm HLL}=
 \begin{cases}
 F_L,&S_L\ge0,\\
 \dfrac{S_RF_L-S_LF_R+S_LS_R(Q_R-Q_L)}{S_R-S_L},&S_L<0<S_R,\\
 F_R,&S_R\le0.
 \end{cases}
\end{equation}
The cell residual is the divergence of these ALE fluxes plus $2Q$, as in
\cref{eq:ss-euler}.  Density and pressure positivity are checked at every line
search trial.

\subsection{Incoming-characteristic boundary projection}

Let $u_n$ and $u_t$ be normal and tangential velocities and define the relative
normal velocity $w=u_n-w_n$.  The acoustic invariants and entropy variable are
\begin{equation}
 J_-=w-\frac{2c}{\gamma-1},
 \qquad
 J_+=w+\frac{2c}{\gamma-1},
 \qquad
 S=\frac{p}{\rho^\gamma}.
\end{equation}
On the subsonic left and top boundaries, the minus acoustic family is incoming
in the adopted orientation.  The ghost state uses $J_-$ from the matched
outer target and $J_+$ from the interior.  On the left boundary, entropy and
tangential velocity remain interior data; on the top subsonic segment they are
also supplied by the matched target.  The rigid wall reflects normal momentum,
and the right boundary uses the exact upstream state.  Characteristic speeds
are always computed relative to the self-similar boundary velocity; using
laboratory velocities changes the boundary classification.

\subsection{Hunter--Tesdall outer target}

Introduce parabolic coordinates
\begin{equation}
 r=\xi+\frac{\eta^2}{4},
 \qquad q=\max(1-r,0).
\end{equation}
The linearised reflected-wave solution used on the subsonic boundary is
specified by
\begin{equation}
 \varphi_r=q+\frac{1}{\pi}
 \operatorname{atan2}\left(2a\sqrt q,
 q+\frac{\eta^2}{4}-a^2\right),
\end{equation}
and
\begin{equation}
 \varphi(r,\eta)=\varphi(1,\eta)-\frac12q^2
 -\frac1\pi\int_0^q
 \operatorname{atan2}\left(2a\sqrt s,
 s+\frac{\eta^2}{4}-a^2\right)\dd s.
\end{equation}
Then
\begin{equation}
 u=\varphi_r+r,
 \qquad
 v=\varphi_\eta+\frac\eta2u.
\end{equation}
These canonical perturbations are lifted linearly about the exact
finite-strength post-incident state.  The lift preserves that state when the
canonical data equal the post-shock canonical velocity and adds only the
leading acoustic perturbation; it does not treat the linearised potential as
an exact nonlinear Bernoulli solution.

\subsection{Sparse coloured Newton solve}

The MUSCL stencil has radius two along each coordinate line.  A $5\times5$
cell colouring, repeated for four conservative components, therefore recovers
the exact sparse Jacobian with 100 batched automatic-differentiation
Jacobian--vector products.  Each regularised Newton step solves
\begin{equation}
 (J+\sigma I)\Delta q=-R
\end{equation}
by sparse LU with fill-reducing ordering.  Backtracking requires residual
reduction and positive density and pressure.  The regularisation is increased
on factorisation or line-search failure and decreased after successful steps.
Every state used in the paper has RMS nonlinear residual at or below
$2\times10^{-6}$.

\section{Boundary homotopy, strength continuation, and conservative remapping}
\label{app:continuation}

The matched branch is reached from the retained incident-boundary solution by
the homotopy
\begin{equation}
 B_\beta=(1-\beta)B_{\rm inc}+\beta B_{\rm char},
 \qquad 0\le\beta\le1,
\end{equation}
where $B_{\rm char}$ denotes the incoming-characteristic projection of the
Hunter--Tesdall target.  Small $\beta$ increments are used until the fully
matched state is obtained.  Strength continuation then proceeds from larger
to smaller $\mu$, using the previous converged state as a predictor.

When the grid changes, conservative overlap remapping is used.  Let
$O^x_{Ii}$ and $O^y_{Jj}$ be the lengths of intersection between new and old
cell intervals.  For each conservative component,
\begin{equation}
 Q^{\rm new}_{JI}=
 \frac{1}{\Delta x_I^{\rm new}\Delta y_J^{\rm new}}
 \sum_{i,j}O^x_{Ii}O^y_{Jj}Q^{\rm old}_{ji}.
 \label{eq:app-remap}
\end{equation}
Thus the domain integral of mass, both momenta, and energy is preserved up to
roundoff.  Each continuation run stores an intermediate checkpoint after every
accepted Newton step; no under-converged state enters the qualification tables.

\section{Euler front extraction and topology signature}
\label{app:front-topology}

\subsection{RH-constrained fitted front}

At each row, a compression ridge is located in a Gaussian window following the
exact incident line.  A soft maximum produces a differentiable row position,
and the principal graph is fitted with the same hinge form as the
potential-flow front.  States are sampled on both sides along the physical
normal.  The Euler Rankine--Hugoniot residual uses the physical self-similar
flux and is normalised by the local state jump.  A branch-continuous initial
guess prevents the optimiser from moving to a nearby secondary ridge.

The 12-member family uses
$d_n\in\{0.075,0.0875,0.1000,0.1125\}$ and
$w_{RH}\in\{0.75,1.0,1.5\}$.  The mean is the reported trajectory and the
standard deviation is a functional-definition diagnostic.  It is not divided
by $\sqrt{12}$ and is not interpreted as sampling noise.

\subsection{Conservative-equivalent front diagnostic}

For a captured scalar profile $q_i$ connecting plateau values $q_L$ and
$q_R$, define the local mixture fraction
\begin{equation}
 f_i=\operatorname{clip}\left(\frac{q_i-q_R}{q_L-q_R},0,1\right).
\end{equation}
The equivalent sharp-step position is
\begin{equation}
 x_s=x_{L-1/2}+\Delta x\sum_i f_i.
 \label{eq:app-equivalent-front}
\end{equation}
This construction conserves the integrated excess relative to the right
state.  It is applied independently to density, pressure, and canonical
streamwise velocity; the row position is their median and their spread is a
local extraction diagnostic.  This front locator is less sensitive than a
maximum-gradient point to the internal shape of a smeared shock and is
retained as an independent audit, not substituted post hoc to tune a
coefficient.

\subsection{Topology signature}

A state is accepted only if the following signature is continuous under
strength and grid continuation:
\begin{enumerate}[label=(\alph*)]
 \item the selected ridge connects to the incident shock and is the lowest
       RH-compatible principal compression branch;
 \item the number of connected local pseudo-subsonic components is unchanged;
 \item wall and sonic-set intersection counts are unchanged;
 \item the ordering of the principal and secondary ridges in $\eta$ is
       preserved;
 \item the attachment coordinates vary continuously; and
 \item the RH defect and fit-family spread remain within the qualified range.
\end{enumerate}
A small nonlinear residual alone is therefore insufficient: a converged state
on the wrong compression branch is rejected.

\section{Asymptotic-preserving grid design and coupled extrapolation}
\label{app:ap-design}

The weak-shock map compresses the physical transverse coordinate by
$\sqrt\mu$.  In the modified equation of a split shock-capturing method, the
transverse acoustic dissipation, after conversion to canonical variables, is
controlled by
\begin{equation}
 \rho_{\rm num}=\frac{h_\eta}{\sqrt\mu}.
\end{equation}
Thus a fixed canonical grid does not define a uniform weak-shock limit:
$\rho_{\rm num}$ increases as $\mu\to0$.  The numerical limit is instead taken
in two stages: first $\mu\to0$ at approximately fixed $\rho_{\rm num}$, then
$\rho_{\rm num}\to0$.

\begin{longtable}{lrrrrr}
\caption{All 21 Euler states used for the leading coupled-limit analysis.}
\label{tab:app-euler-states}\\
\toprule
Slice & $\mu$ & $N_\xi$ & $N_\eta$ & $\rho_{\rm num}$ & mean $g$\\
\midrule
\endfirsthead
\toprule
Slice & $\mu$ & $N_\xi$ & $N_\eta$ & $\rho_{\rm num}$ & mean $g$\\
\midrule
\endhead
coarse & 0.100000 & 68 & 40 & 0.118585 & 0.459960\\
medium & 0.100000 & 84 & 48 & 0.098821 & 0.460076\\
fine & 0.100000 & 112 & 64 & 0.074116 & 0.463658\\
coarse & 0.075000 & 80 & 46 & 0.119070 & 0.467062\\
medium & 0.075000 & 96 & 56 & 0.097808 & 0.467272\\
fine & 0.075000 & 132 & 76 & 0.072069 & 0.467391\\
coarse & 0.050000 & 96 & 56 & 0.119789 & 0.471088\\
medium & 0.050000 & 116 & 68 & 0.098650 & 0.470008\\
fine & 0.050000 & 160 & 92 & 0.072915 & 0.471916\\
coarse & 0.040000 & 108 & 64 & 0.117188 & 0.471846\\
medium & 0.040000 & 128 & 76 & 0.098684 & 0.472981\\
fine & 0.040000 & 172 & 100 & 0.075000 & 0.475964\\
finest & 0.040000 & 220 & 128 & 0.058594 & 0.478343\\
coarse & 0.033333 & 120 & 70 & 0.117369 & 0.472727\\
medium & 0.033333 & 140 & 82 & 0.100193 & 0.474920\\
fine & 0.033333 & 192 & 112 & 0.073356 & 0.478014\\
finest & 0.033333 & 240 & 140 & 0.058685 & 0.480705\\
coarse & 0.025000 & 136 & 80 & 0.118585 & 0.476717\\
medium & 0.025000 & 164 & 96 & 0.098821 & 0.476437\\
fine & 0.025000 & 220 & 128 & 0.074116 & 0.480546\\
finest & 0.025000 & 256 & 148 & 0.064100 & 0.481965\\
\bottomrule
\end{longtable}

For each response $Y$, the maximal retained model is
\begin{equation}
 Y=Y_0+a_1\mu+a_2\mu^2+b_1\rho+b_2\rho\sqrt\mu+b_3\rho^2.
\end{equation}
Restricted models remove one or more curvature or interaction terms.  For
$n$ data and $k$ fitted parameters,
\begin{equation}
 \mathrm{AICc}=n\log(\mathrm{RSS}/n)+2k+
 \frac{2k(k+1)}{n-k-1}.
\end{equation}
Model assessment also includes pointwise leave-one-out error,
leave-one-resolution-slice-out error, restriction to $\mu\le0.05$, and all 12
front functionals.  The leading uncertainty envelope combines variation over
these defensible choices; no single polynomial is selected because it happens
to lie closest to the potential-flow value.

\section{Same-strength phase brackets and cubic secants}
\label{app:phase-details}

The captured-shock subcell phase is measured with three nearby grids at each
of two fixed strengths and three stretched-resolution levels.  The brackets
are
\begin{longtable}{lrrrrr}
\caption{Three-grid phase-bracket summaries.  $G$ denotes
$\theta_{\rm phys}/\sqrt\mu$.}
\label{tab:app-phase-brackets}\\
\toprule
Level & $\mu$ & mean $\rho$ & mean $g$ & half-range $g$ & mean $G$\\
\midrule
\endfirsthead
\toprule
Level & $\mu$ & mean $\rho$ & mean $g$ & half-range $g$ & mean $G$\\
\midrule
\endhead
fine & 0.025 & 0.074128 & 0.480366 & 0.001129 & 0.234440\\
medium & 0.025 & 0.098850 & 0.477071 & 0.001302 & 0.232335\\
coarse & 0.025 & 0.118635 & 0.476017 & 0.000712 & 0.231664\\
fine & 0.050 & 0.072938 & 0.472851 & 0.000816 & 0.227146\\
medium & 0.050 & 0.098707 & 0.470521 & 0.000853 & 0.225691\\
coarse & 0.050 & 0.119891 & 0.470047 & 0.000867 & 0.225395\\
\bottomrule
\end{longtable}
The phase-averaged secant is
\begin{equation}
 K_{2,\mathrm{sec}}^{\Eul}(\rho)=
 \frac{G^{\Eul}(0.05,\rho)-G^{\Eul}(0.025,\rho)}{0.025}.
\end{equation}
If $e_{0.05}$ and $e_{0.025}$ are the three-grid half-ranges, the conservative
phase envelope is
\begin{equation}
 e_K=\frac{\sqrt{e_{0.05}^2+e_{0.025}^2}}{0.025}.
\end{equation}
The resulting values are reported in \cref{tab:phase-secants}.  They are all
compatible with the potential-flow diagnostic at finite $\rho$, but their
central values drift with refinement.  A constant model ignores the drift, a
linear model extrapolates outside the sampled range, and a quadratic model has
no predictive redundancy with only three resolution levels.

\begin{figure}[p]
 \centering
 \includegraphics[width=0.82\linewidth]{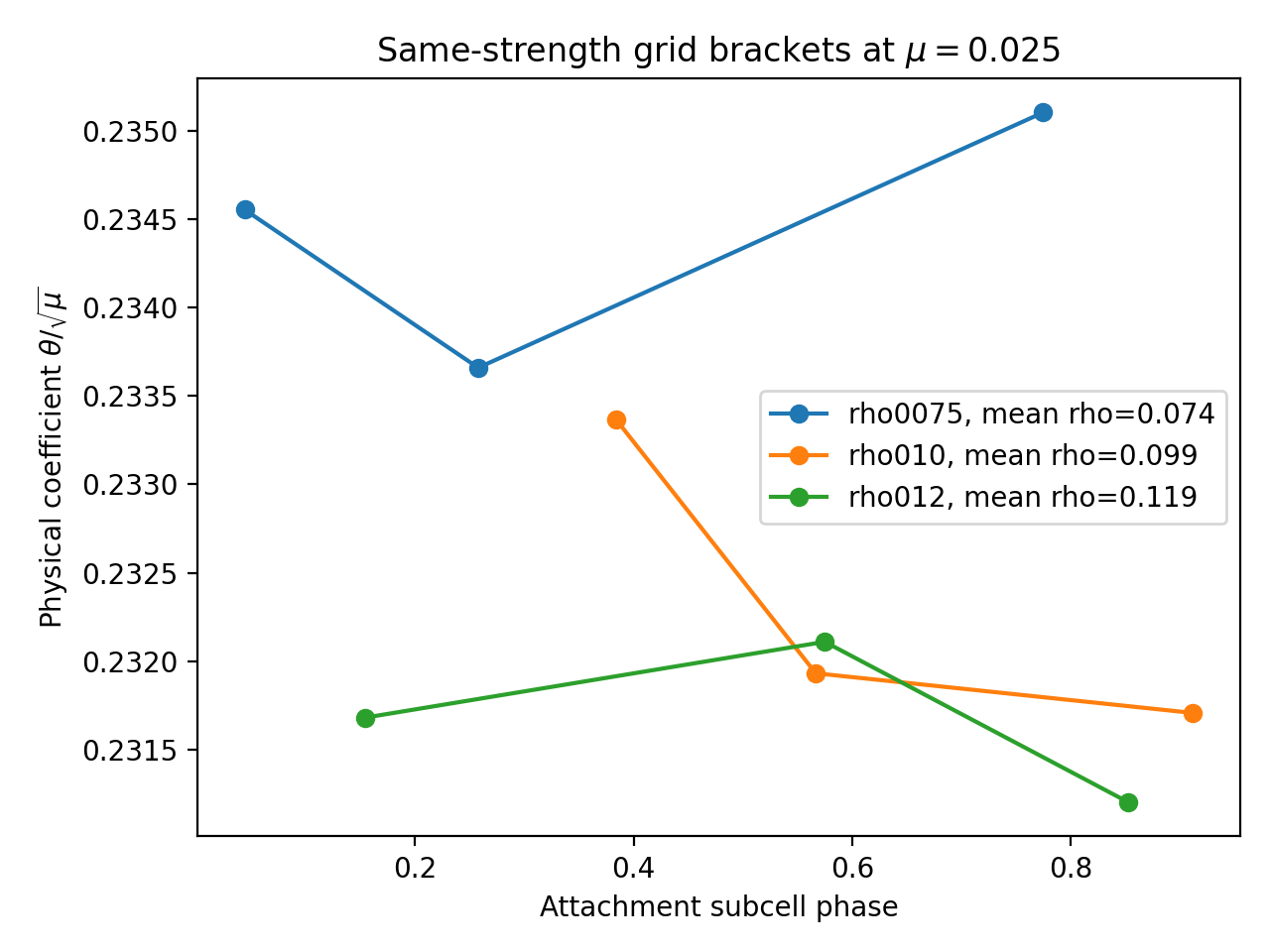}
 \caption{Same-strength three-grid brackets at $\mu=0.025$.  Variation with
 subcell phase is comparable to the desired cubic signal.}
 \label{fig:app-phase-0025}
\end{figure}

\begin{figure}[p]
 \centering
 \includegraphics[width=0.82\linewidth]{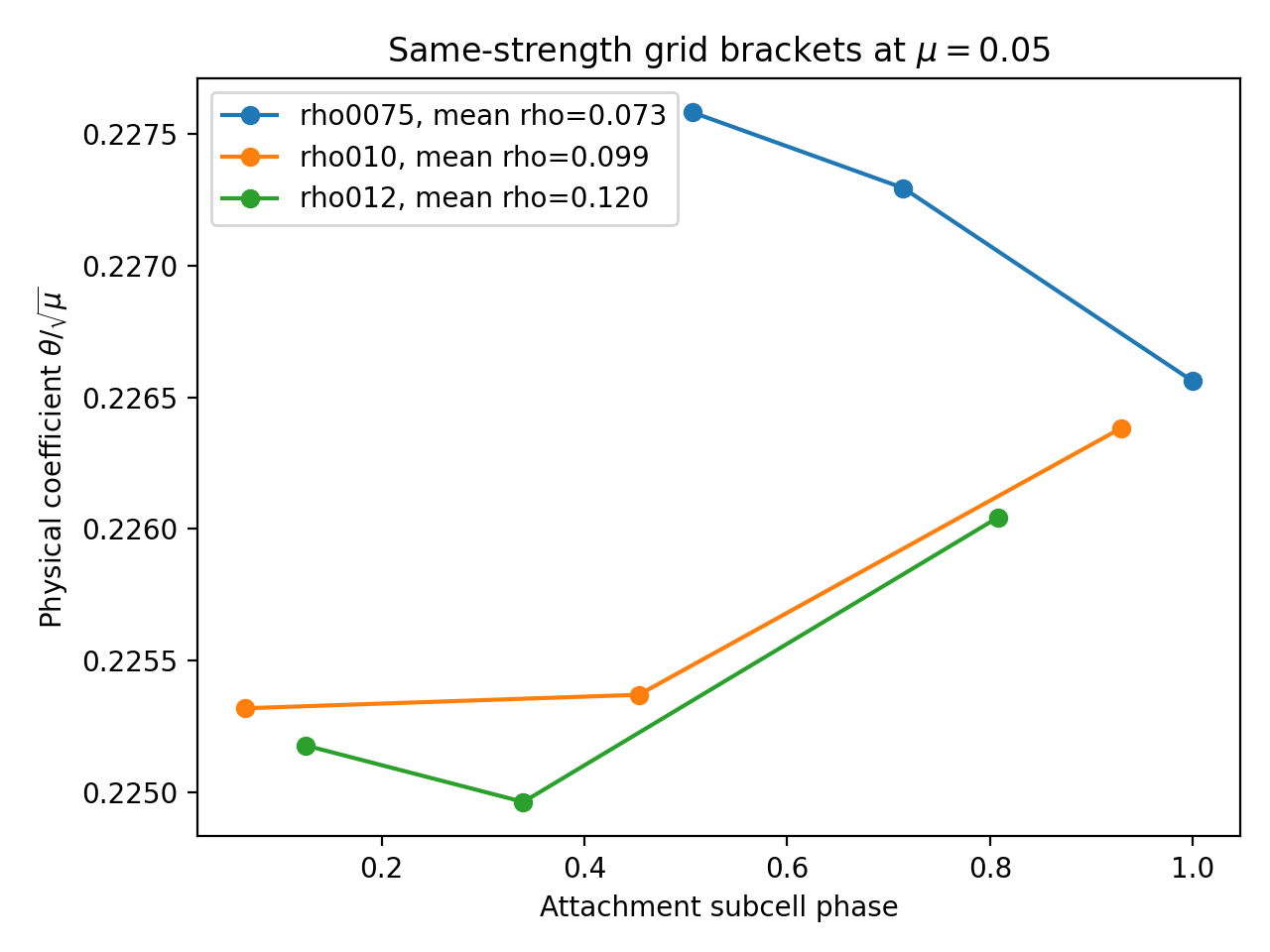}
 \caption{Same-strength three-grid brackets at $\mu=0.05$.  The same phase
 structure persists at the stronger member of the cubic secant.}
 \label{fig:app-phase-005}
\end{figure}

As an additional diagnostic, the nine grids at each strength were fitted to
restricted forms of
\begin{equation}
 Y(\rho,\phi)=Y_0+c_1\rho+c_2\rho^2
 +A\sin(2\pi\phi)+B\cos(2\pi\phi),
\end{equation}
with optional $\rho$-dependent phase amplitude.  The resulting continuum
secants are strongly model-dependent and the curved models are ill-conditioned.
This confirms that phase modelling does not create a uniquely identified
$\rho\to0$ cubic coefficient.

\section{Numerical acceptance, uncertainty, and reproducibility}
\label{app:reproducibility}

The error sources are tracked separately:
\begin{equation}
 \varepsilon_{\rm total}
 =\varepsilon_{\rm residual}+\varepsilon_{\rm grid}
 +\varepsilon_{\rm phase}+\varepsilon_{\rm functional}
 +\varepsilon_{\rm model}+\varepsilon_{\rm branch}.
\end{equation}
This is a bookkeeping identity, not an assumption that the terms are
independent random variables.  In particular:
\begin{itemize}
 \item residual error is controlled by nonlinear stopping criteria and
       tangent/adjoint residuals;
 \item grid error is assessed across the AMR hierarchy or the $\rho$ slices;
 \item phase error is measured by same-strength neighbouring-grid brackets;
 \item functional error is the spread over the fixed 12-member front family;
 \item model error is the variation over defensible coupled extrapolations;
 \item branch error is controlled by the topology signature and continuation.
\end{itemize}
No quoted interval is a statistical confidence interval.

The complete acceptance hierarchy is:
\begin{enumerate}[label=(\arabic*)]
 \item nonlinear residual convergence for every retained state;
 \item independent residual and symbolic parent checks;
 \item Jacobian/transpose dot-product tests;
 \item centred differences of explicit parameter derivatives;
 \item tangent, adjoint, and complete re-solve agreement;
 \item fixed dimensional front offsets and common fit family;
 \item RH and secondary-ridge audits;
 \item topology continuity under grid and strength continuation;
 \item coupled $\mu$--$\rho$ model and cross-validation tests;
 \item same-strength phase brackets at multiple $\rho$ levels;
 \item exact conversion from $(\xi_T,\eta_T)$ to the physical angle; and
 \item explicit distinction between fixed-$a$ and fixed-$\lambda$ paths.
\end{enumerate}

The accompanying reproducibility archive contains the original retained V13
package, reconstructed matched-boundary source tree, automated tests,
continuation scripts, nonlinear checkpoints, 21-state leading dataset,
18-state phase-bracket dataset, all 12-member front fits, figure-generation
scripts, manuscript source, file inventory, and SHA-256 checksums.  The main
paper can therefore be reproduced from compact CSV/JSON inputs, while the full
archive permits rerunning the nonlinear calculations.

\end{document}